\documentclass[twocolumn,showpacs,superscriptaddress,twoside,prl]{revtex4}
\usepackage{mathrsfs, doi, hyperref,bm}
\usepackage{amssymb, amsbsy, amsmath, latexsym, dsfont, array, layout, graphicx,mathrsfs,braket,amsfonts,amsthm, amssymb,graphicx,subfigure, natbib, youngtab,color,xcolor,hyperref}
\newcommand{\one}{\mathds{1}}

\begin{document}

\title{
	Differential Evolution for Many-Particle Adaptive Quantum Metrology
	}
\author{Neil B. Lovett}
\affiliation{Institute for Quantum Science and Technology, University of Calgary, Alberta T2N 1N4, Canada}
\author{C\'ecile Crosnier}
\affiliation{Institute for Quantum Science and Technology, University of Calgary, Alberta T2N 1N4, Canada}
\affiliation{D\'epartement de Physique, \'Ecole Normale Sup\'erieure de Cachan, 94230 Cachan, France}
\author{Mart\'i Perarnau-Llobet}
\affiliation{Institute for Quantum Science and Technology, University of Calgary, Alberta T2N 1N4, Canada}
\affiliation{Institute for Theoretical Physics, University of Amsterdam,
	1090 GL Amsterdam, The Netherlands}
\affiliation{ICFO-Institut de Ciencies Fotoniques, Mediterranean Technology Park, 08860 Castelldefels (Barcelona), Spain}
\author{Barry C. Sanders}
\email{sandersb@ucalgary.ca}
\affiliation{Institute for Quantum Science and Technology, University of Calgary, Alberta T2N 1N4, Canada}

\begin{abstract}
We devise powerful algorithms based on differential evolution
for adaptive many-particle quantum metrology.
Our new approach delivers adaptive quantum metrology policies
for feedback control that are orders-of-magnitude more efficient
and surpass the few-dozen-particle limitation arising in methods based on
particle-swarm optimization.
We apply our method to the binary-decision-tree model
for quantum-enhanced phase estimation as well as to a new problem: 
a decision tree for adaptive estimation of the unknown bias of a quantum coin
in a quantum walk and show how this latter case can be realized experimentally.
\end{abstract}

\date{\today}
\pacs{03.67.-a, 03.67.Ac}

\maketitle

\emph{Quantum-enhanced metrology} (QEM)
aims to achieve single-shot parameter estimation
for Hamiltonian-generated evolution of~$N$ particles with a degree of imprecision~$\Delta_N$
(e.g., standard deviation) exceeding the
semiclassical limit (or ``standard quantum limit'').
This limit is due to
particle partition noise (vacuum fluctuations)~\cite{wiseman10a}
and ultimately restricts the precision of  clocks \cite{ye08a},
gravitational wave detection~\cite{abbott09a} 
and adaptive Hamiltonian identification~\cite{SKO04}.
Mathematically, $\Delta_N\in O\left(N^{-\wp}\right)$
with~$N$ the number of particles in the probe ``pulse''
(our term for a collection of particles, e.g., photons)
and~$\wp=1/2$ ($\wp=1$) in the semiclassical (ultimate) precision limit~\cite{Luo00,ZPK10,ZPK11E}.
The objective of single-shot QEM is to attain precision exceeding~$\wp=1/2$
and reaching as close as possible to~$\wp=1$ for a single ``pulse'',
as opposed to tomography where many ``pulses'' could be used.
Two common QEM strategies inject quantum-resource-laden (e.g., entangled) input states
(i)~followed by multi-particle joint measurement or
(ii)~our focus: \emph{adaptive} QEM (AQEM),
which employs only local measurements
each followed by optimal control of system parameters in order to
extract maximal information about unknown parameters~\cite{giovannetti11a}.

\begin{figure}
\includegraphics[width=\columnwidth]{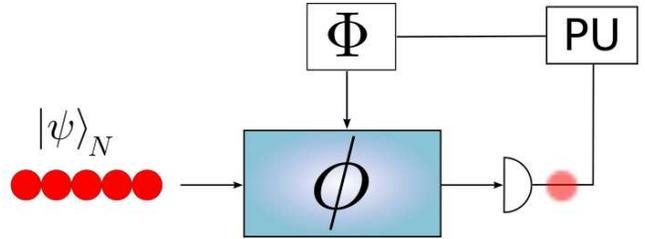}
\caption{
	(Color online) A set of (possibly entangled) particles, solid (red) circles on the LHS, are injected into a system with unknown parameter~$\phi$.
	Information from sequential measurements on each outgoing particle, faded (red) circle on the RHS, is fed to a processing unit (PU)
	to modify a control parameter~$\Phi$ to enhance the precision of estimating~$\phi$.
	Machine learning is used on training sets to find a suitable decision-tree-based algorithm for the PU
	so that single-shot estimates of~$\phi$ beat the semiclassical measurement limit.
	}
\label{fig:scheme}
\end{figure}

Finding effective adaptive-feedback procedures (known as ``policies'' in machine learning) is 
typically intractable
but facilitated by decision-tree learning~\cite{HS10,HS11}.
Here we report three major new advances in AQEM
enabled via our introduction of differential-evolution (DE)~\cite{storn97a}
decision-tree learning to AQEM:
(a)~surpassing the few-dozen-particle limit in previous interferometric-phase-estimation studies~\cite{HS10,HS11}
explained and depicted in Fig.~\ref{fig:scheme};
(b)~advancing beyond the binary decision tree for quantum-walk coin-bias parameter estimation;
and (c)~showing how our learning algorithm can be used in optical quantum experiments with current technology~\cite{schreiber09a}.
We introduce DE as a tool for AQEM because of its known superiority over 
the \emph{Particle} Swarm Optimization (PSO)
machine-learning algorithm~\cite{kennedy95a}
for many optimization problems, especially for high-dimensional search spaces~\cite{vesterstrom04a, das11a},
hence appropriate for AQEM.
(We immediately see an ambiguity of terminology:
the \emph{particle} traversing the interferometer
is different from the machine-learning \emph{particle},
which is a test function in a search space.
As \emph{particle} is a common term in both quantum physics and machine learning,
we will use the term in both ways and make the term clear through context.)
Whereas previous work~\cite{HS10,HS11}
demonstrated that swarm
(collection of particles)
intelligence yields AQEM algorithms superior to algorithms so far devised by sentient beings (i.e., humans),
our use of an evolutionary algorithm here goes beyond an in-principle demonstration of 
artificial intelligence for AQEM towards a realistic approach to devising algorithms for many-particle systems.

PSO is inspired by a social-behavior model comprising~$\Xi$ individual ``particles'' 
stochastically searching a vector space
punctuated by~$\Upsilon$ iterations of mutual communication and 
collective-intelligence decisions to circumvent local-minima traps.
The PSO AQEM algorithm employs a highly effective logarithmic-search heuristic
to devise policies for single-shot AQEM interferometric 
phase estimation~\cite{HS10,HS11}.
Here a policy is defined to be a procedure that an ``agent'',
representing the feedback loop,
adopts given a set of measurement results
for a subset of particles in the output pulse.
A good policy, 
namely, a policy that beats the semiclassical measurement limit,
was previously obtained with a computational-space overhead~$O(N)$
accompanied by a run-time cost~$O(N^6)$~\cite{HS11}.
Here we show that this aforementioned PSO-based algorithm breaks down
for just dozens of particles,
but we remedy this limitation here by switching the learning algorithm from PSO to DE 
(which we show dramatically speeds up the simulation run-time)
but pay a time-cost slight penalty, namely~$O(N^7)$
instead of the previous~$O(N^6)$,
thereby surpassing the previous maximum-number-of-particles barrier to devising policies.

Our employment of a DE AQEM algorithm also enables us to go beyond the
restrictive binary-outcome measurement model for two-output-port interferometry.
We introduce an example of a single-shot AQEM problem
with a higher number of possible measurement outcomes hence a larger $d-$ary tree.
Specifically we now solve the harder case of a discrete-time quantum walk 
with~$N$ walkers (effectively a ``pulse'' of walkers)
and a quantum-coin operator that has an unknown bias.
The AQEM objective is to ascertain the quantum coin's bias
with an imprecision that scales
better than semiclassical limit $\wp=1/2$.
As a position measurement of the walker at time~$t$
yields an outcome in $\{-t,\dots,t\}$,
the resultant decision tree is $d$-ary for $d=2t+1$.
Our strategy is to replace the $d\propto t$ tree by a 
quaternary ($d=4)$ decision tree and show the effectiveness of DE
for finding a policy that beats the semiclassical limit.
Furthermore we propose a feasible quantum optical quantum-walk experiment
that can attain the semiclassical limit and potentially beat it by exploiting entangled photons.

Let us now establish a mathematically rigorous AQEM model.
In the lossless, decoherence-free case,
an $N$-particle input ``pulse'' state
$|\psi\rangle\in\otimes_{i=1}^{N} \mathscr{H}_i$
is acted on sequentially particle-by-particle
by a device with unknown parameter~$\phi$,
which could be a multicomponent vector~$\bm\phi$,
according to
$\mathcal{D}(\phi;\Phi_i):\mathscr{H}_i\to\mathscr{H}_i$
with~$\Phi_i$ a control parameter
(possible a multi-component vector as well)
that is modified according to the measurement history on previous particles.
Each $\mathcal{D}$-transformed particle is measured according to 
$\mathcal{M} : \mathscr{H}_{i} \to \mathcal{O}_i$
for~$\mathcal{O}$ a set of measurement outcomes.
For the interferometer $\mathcal{O}_i=\{0,1\}$;
for the quantum walk $\mathcal{O}_i=\{-i, \hdots, i\}$.
Although~$\mathcal{D}$ generically has a $2^N\times 2^N$ representation,
this reduces to $N\times N$ for a permutationally-symmetric input state~$|\psi\rangle$~\cite{HS11a}.
The sequence~$\bm{\Phi}=\{\Phi_i\}$ is the policy for controlling the interferometer 
in order to extract a measurement
of~$\phi$ with low imprecision~$\Delta_N$.
Our aim is to devise an efficient algorithm that delivers a fit policy~$\bm{\Phi}$
such that~$\Delta_N$ scales better than~$\wp=1/2$,
and each policy is a test function, or particle, in the machine-learning procedure.

Our policy-devising algorithm, which uses machine learning, has the following inputs:
number of particles~$N$,
permutationally-symmetric input state $|\psi\rangle\in P\mathbb{C}^{N+1}$,
a prior probability distribution~$\mathcal{P}$ for the unknown system parameter $\phi$
(typically uniform),
the device operator $\mathcal{D}(\phi)\in\mathbb{C}^{N+1}\times\mathbb{C}^{N+1}$, 
the set of projectors ~$\bm{\Pi_i}$ for each $i^\text{th}$ particle
($|j\rangle\langle j|$ for $j\in\{0,1\}$ in the interferometer case
and $j\in\{-i,\dots,i\}$ for the quantum-walk case),
an integer~$l$ to determine which machine-learning algorithm to use such as PSO or DE,
the number~$\Xi$ of particles, or ``chromosomes'' in DE parlance,
number~$\Upsilon$ of iterations,
the fitness functional $\mathcal{F}$ that assesses the precision guaranteed by executing the policy,
and the maximum number~$\Omega$ of repetitions the machine-learning algorithm
is permitted to run before aborting.
From the multitude of available machine learning techniques,
we compare the two powerful cases of PSO and DE
to devise policies that deliver AQEM parameter estimation.

\begin{figure}
	(a)\includegraphics[width=0.42\columnwidth]{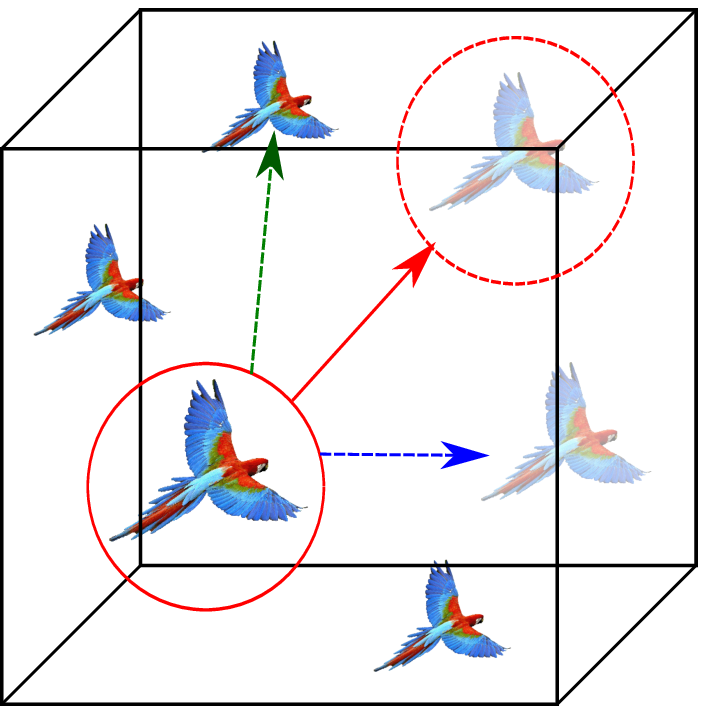}
	(b)\includegraphics[width=0.42\columnwidth]{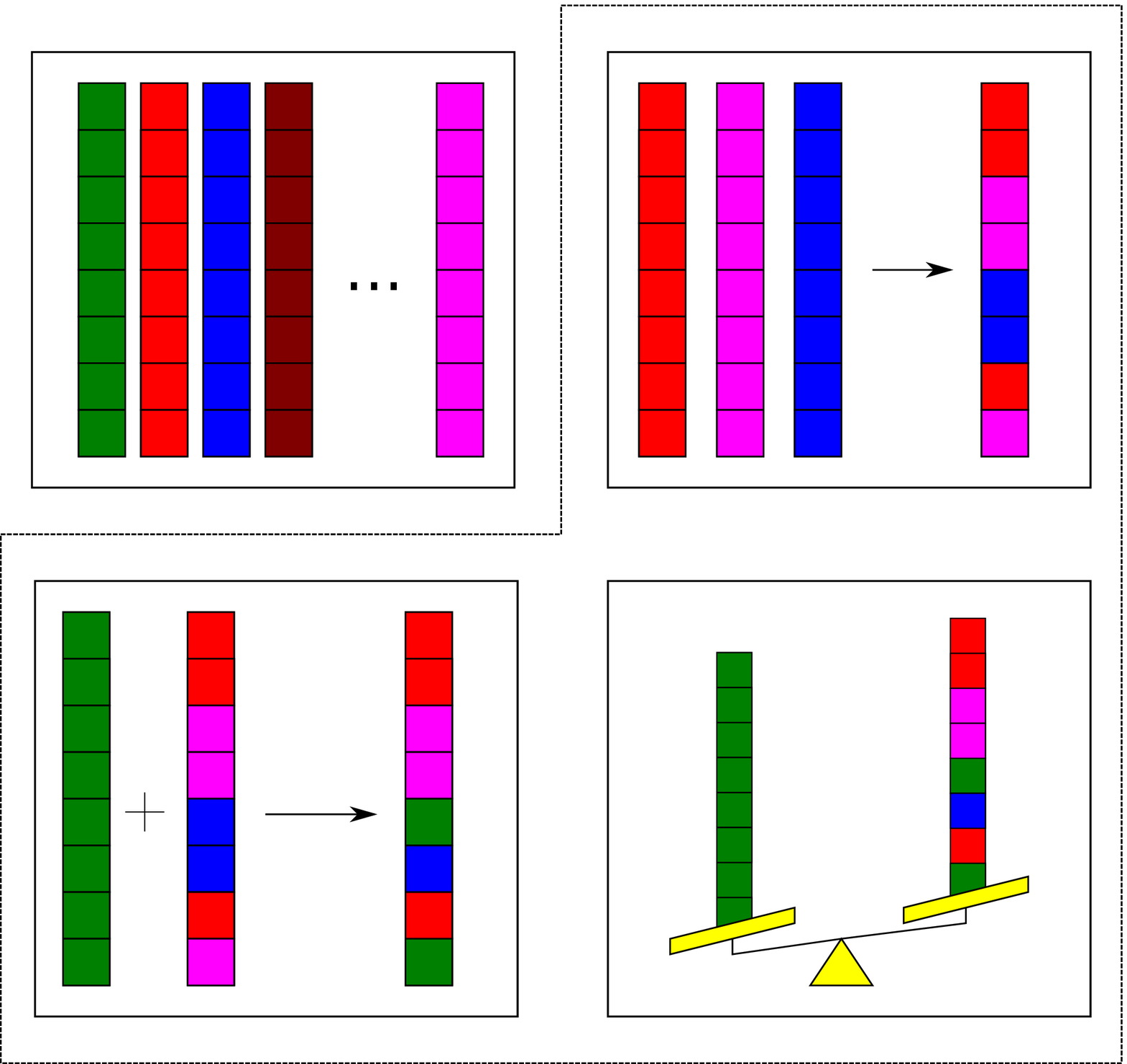}
	\caption{(Color online)~
	(a)~Pictorial representation of the PSO algorithm.
	Each particle, represented here by a bird, stores its current position (solid (red) circle) in the search space (actually parameters
	of the decision tree),
	its best previous position (top left (green arrow)) and the best neighbour position (bottom right (blue arrow)).
	Each bird undergoes simultaneous velocity vector updates (to the dashed (red) circle) according to three terms:
	an inertial term limiting the change in velocity plus two terms that rescales and redirects the velocity 
	to its own personal best and to the best bird in the neighbourhood, respectively.
	(b)~Pictorial representation of the DE algorithm.
	Each chromosome is a vertical block (of decision-tree parameters)
	and initialized to random values in the search space (top left).
	For each chromosome, three random chromosomes are chosen to be parents
	of a donor chromosome comprising random data from each parent (top right).
	This donor chromosome is crossed randomly with the original chromosome to create a trial chromosome (bottom left),
	which is compared with the original (bottom right), and the fitter chromosome is retained for the next iteration.
	The dashed line represents a single iteration of the differential algorithm.
	}
\label{fig:machine}
\end{figure}

The PSO algorithm is based on having multiple particles undergoing independent stochastic 
searches interrupted by periodic iterations of communication between overlapping logarithmic-sized 
neighborhoods of particles that tend to steer these particles 
depicted in Fig.~\ref{fig:machine}(a), 
towards superior policy regions of the vector space.
Similarly DE also employs multiple policies undergoing independent stochastic searches.
Instead of interruptions by rounds of communication and steering,
DE is interrupted by a cross-over breeding between the original chromosome and a hybrid of three randomly 
chosen chromosomes from the remaining set of policies.
The fittest of the original vs the cross-over of the original with the hybrid
is retained for the next round;
see Fig.~\ref{fig:machine}(b).

Algorithmic specific inputs for PSO are 
exploration weight $\alpha$, exploitation weight $\beta$, velocity clamping $\nu$ and inertial weight $\omega$.
For DE, the algorithmic inputs are mutation scaling $\mu$ and cross-over rate $\gamma$.
In order to perform a fair comparison between PSO-- and DE--based adaptive policy-devising algorithms,
we ensure that all common input parameters are identical and parameters specific to PSO or DE are optimized.
Now we consider how to make the policy-devising algorithm efficient and also determine the space and time complexities.
We reduce the space complexity by employing a logarithmic-search heuristic 
that parametrizes the decision tree only by its depth,
and the depth equals $N$ implying a
space cost $O(N)$~\cite{HS10}.

\begin{figure}
	(a) \includegraphics[width=0.42\columnwidth]{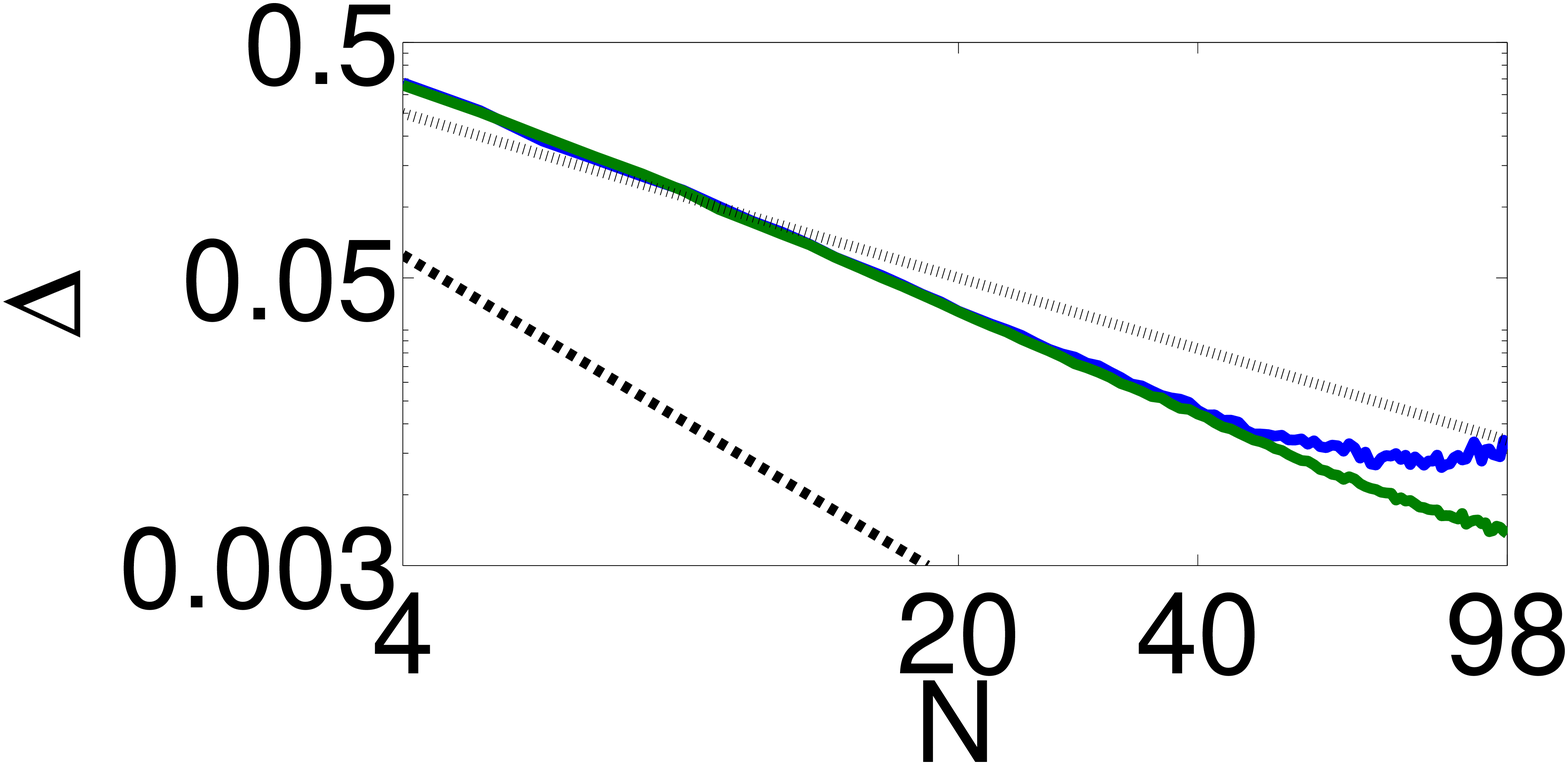}
	(b) \includegraphics[width=0.42\columnwidth]{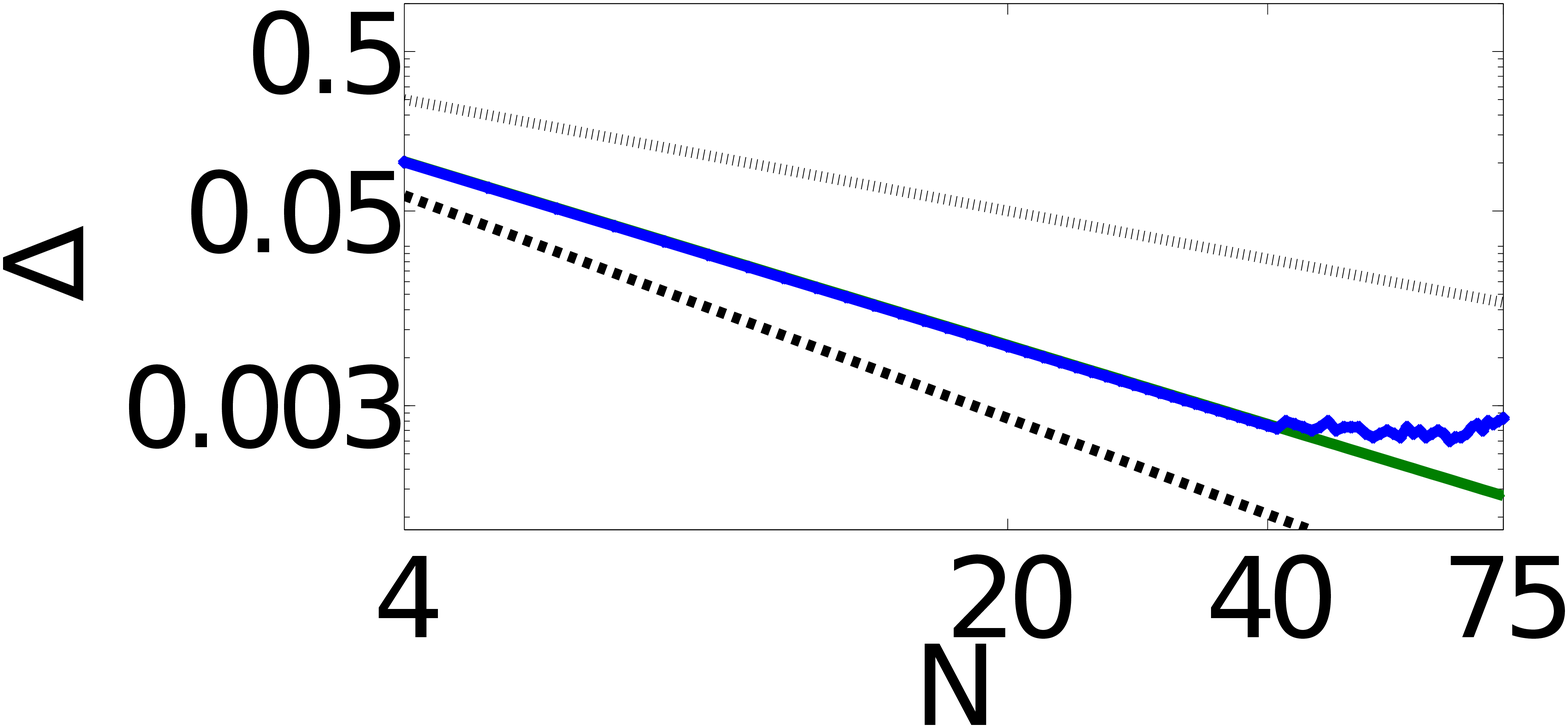}
\caption{
	(Color online) Imprecision~$\Delta$ of (a)~the interferometric phase and
	(b)~the quantum-walk coin bias
	for the semiclassical (uppermost dotted line)
	and ultimate quantum-limited (lowest dashed line) cases
	DE (straight green middle line)), PSO (blue middle line that tilts upward for large~$N$).
	The PSO-- and DE--based plots each required (a)~315
	and (b)~403 CPU-hours on a cluster of 100 parallel cores each running at 2.66GHz
	and show $\wp=0.74$.
	}
\label{fig:phase}
\end{figure}

We develop heuristics to ensure a polynomial time cost:
(i)~simulating the interferometer for a single $N$-particle pulse is~$O(N^2)$~\cite{HS11a}$\S$4.2;
(ii)~iterating the search steps~$\Upsilon\in O(N)$,
which is higher than previous studies that set $\Upsilon\in O(1)$ but enables breaking the few-dozen-particle limit in that work~\cite{HS10,HS11} for the the DE case
but \emph{not} for the PSO case as shown in Fig.~\ref{fig:phase}(a);
(iii)~assessing a candidate policy from~$K\in O(N^2)$ samplings~\cite{HS11};
(iv)~repeating for each of the~$\Xi\in O(N)$ particles; and
(v)~constructing a starting distribution from the~$(N-1)$-particle policy 
with concomitant time cost $\in O(N)$.
The $N$-particle policy imprecision~$\Delta_N$,
determined from the preceding fittest $(N-1)$-particle policy
has a ratio $\Delta_N/\Delta_{N-1}=1-1/N$,
which necessitates $\Omega\in O(N)$ repetitions of the algorithm. 

The adaptive interferometric phase-estimation algorithm commenced
with initial (unnormalized) multi-particle two-mode entangled state~\cite{HS10,HS11}
$\sum_{n,k=0}^N\sin(\frac{k+1}{N+2}\pi)
	\text{e}^{i\pi (k-n)/2}d_{n-N/2,k-N/2}^{N/2}(\frac{\pi}{2})|n,N-n\rangle$
with~$d^\bullet_{\bullet,\bullet}$ a reduced rotation matrix element~\cite{BS68},
and fitness function for the phase-error distribution as
$|\int_{-\pi}^{\pi}P(\zeta)e^{i\zeta}d\zeta|$
with~$\zeta$ the absolute difference between inferred and correct phase in a training set.
We see in Fig.~\ref{fig:phase}(a) that the adaptive interferometric-phase-estimation policies 
found using DE surpass those found using PSO 
in that they maintain the power-law scaling (better than semiclassical limit)
past the few-dozen particle-number limit. We are able to simulate up to 98 particles in the input state with no sign of breakdown.
As we use the same space and time resources for the PSO- and DE-based algorithms,
an improvement from simulating up to 45 particles in the former algorithm compared to 98 particles in the latter 
corresponds to a $(\frac{98}{45})^7\approx 232$-fold effective increase in run-time.

As DE is so much more powerful than PSO for adaptive quantum metrology,
we consider solving a significantly more challenging AQEM problem,
specifically estimating the bias~$\phi$ of a quantum-walk coin~\cite{tregenna03a}.
The key challenge is due to the larger number of measurement outcomes
than just two for interferometric phase estimation.
For each walker, the walker-coin basis states at time~$t$ are
$\{|x,c\rangle:x \in \{-t,\hdots,t\},c\in \{ -1,1\}\}$
with dimension~$d_t=2(2t+1)$.
Each quantum-walk step is a sequence of a coin flip
$C(\phi) |c \rangle = \sqrt{\phi}|-1 \rangle + c \sqrt{1-\phi} |1\rangle$
and a conditional walker translation~$S | x, c \rangle = | x+c, c \rangle$.
The step operation $S\left(\one\otimes C\right)$ is repeated~$t$ times.

The procedure to estimate bias~$\phi$ is similar to estimating the interferometric phase in that a single pulse of sequential particles is
injected to a quantum-walk apparatus of duration~$t$ where the particles in this case are quantum walkers.
Unlike the interferometric case where each particle
is equally likely to traverse each of  two available paths,
here the bias causes an unequal split between multiple paths,
in contrast to the classical case where the bias shifts the walker's distribution left or right, 
the quantum-biased coin alters the shape of the distribution.

We assume an initial $N$-walker input,
adapted from the two-walker state~\cite{omar06a},
such that the state is permutationally symmetric,
in order to ensure algorithmic time cost $O(N^2)$ as in the interferometric-phase case.
Furthermore each walker's initial state is symmetrized with respect to the position
around $x=0$.
The position distribution for the walker's reduced state (tracing over the coin state)
becomes increasingly asymmetric due to the bias of the coin, 
and we introduce the skewness of this distribution
(given that the quantum-coin bias alters the distribution shape)
as the fitness parameter in the machine-learning algorithm.
This machine-learning algorithm is part of an AQEM algorithm responsible for
finding a fit feedback policy that determines how much to modify the coin's bias subsequent to each single-particle measurement.

For estimating the coin bias,
we introduce an effective heuristic based on reducing the $d$-ary decision tree to a quaternary ($d=4$) decision tree
and maintaining the logarithmic search heuristic developed for the interferometric-phase case \cite{HS10}.
The reason for $d=4$ begins with recognizing that the quantum walker's position distribution
can be broken up into four regions given by left-outer, left-inner, right-inner and right-outer.
As is well known for the coined quantum walk,
the inner region of the position distribution contains 1/3 of the position probability and is approximately uniform.
The outer region contains the remaining 2/3 of the distribution and is highly peaked~\cite{ambainis01a}.
The skewness of the distribution is expected to show more strongly by comparing the left and right \emph{outer} regions
rather than restricting to the binary case of comparing the entire left and right regions.

We execute the policy-devising algorithm with this $d=4$ heuristic with $O(N)$ space cost and $O(N^7)$ time cost as before.
Figure~\ref{fig:phase}(b) shows imprecision~$\Delta_N$ of policies found using PSO and DE with the semiclassical and ultimate quantum power-law limits for reference,
where~$\Delta_N$ is the imprecision not of~$\phi$ but rather of~$t\phi$ because the biased coin operation
has been executed $t$~times.
Specifically Fig.~\ref{fig:phase}(b) shows power-law scaling
for up to 35 walkers per pulse in the PSO case and fails beyond 35 walkers.
Contrariwise the DE-based adaptive metrology algorithm successfully determines policies that maintain power-law scaling
up to 75 walkers with no sign of power-law breakdown.
The resultant improvement from 35 to 75 walkers by using DE instead of PSO corresponds to a 
$(\frac{75}{35})^7\approx 208$-fold decrease of effective run-time, which is comparable to the speed-up for the interferometric-phase case.

\begin{figure}
\includegraphics[width=\columnwidth]{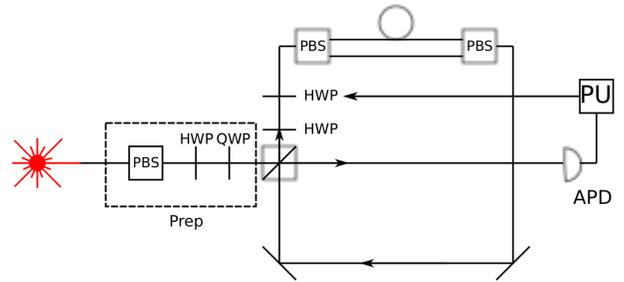}
\caption{
	(Color online)
	Schematic of proposed adaptive quantum metrology experiment to determine the bias of the half-wave plate (HWP)
	A laser source (red star) is attenuated to the single-photon level.
	The field then passes through a polarizing beam splitter (PBS), then a HWP, followed by a quarter-wave plate (QWP)
	to prepare (`Prep') any initial walker state.
	The beam splitter is controlled by an active switch, which determines whether the photon re-enters the loop or is sent to the
	avalanche photo diode (APD) detector (after~$t$ steps).
	Once in the fiber network, the unknown bias coin operation is performed using a half wave plate with unknown angle~$\theta$. 
	The adaptive coin operation is effected by another HWP with~$\theta$ modified by the processing unit (PU)
	subsequent to measurement of the previous walker's time of arrival.
	The photons then pass through a delay loop with PBSs on either side effecting the shift-in-time operation.
	}
\label{fig:optics}
\end{figure}

AQEM for quantum walks is particularly exciting because implementation is feasible
with existing quantum optical quantum-walk experimental techniques~\cite{schreiber09a} as we now show.
In this approach quantum walkers are photons, and the position degree of freedom is replaced by time of arrival.
The coin state corresponds to the polarization state of the photon,
and coin flips are executed using a half-wave plate (HWP),
which transforms the polarization to a superposition of the two polarizations that can be unequally weighted according to 
the angle~$\theta$ of the HWP relative to one of the polarization axes.
Quantum-walk steps are implemented by having the photons circumambulate an optical fiber loop as depicted in Fig.~\ref{fig:optics}.
A 50:50 beamsplitter enables the photon to exit the loop leading to an avalanche photo diode (APD) where the position of the walker is realized temporally as an arrival time.
Thus, there is only a 50\% chance the photon will remain in the fiber loop
and advance to the next step.

In the model we propose, each walker performs~$t$ steps before being measured.
Our modification to existing experiments is shown in Fig.~\ref{fig:optics}.
This modification replaces the 50:50 beamsplitter with an active switch into the detection fiber (as suggested earlier~\cite{schreiber09a}).
This switch allows for a controllable number~$t$ timesteps.
The `biased' coin is achieved using the HWP with an unknown angle~$\theta$. The adaptive coin operation is implemented by another HWP with the angle~$\theta$ controlled by a processing unit programmed with the specific feedback policy found by our algorithm. Our heuristic of grouping the measurement outcomes is accomplished by translating those groupings into arrival time bins.
Thus our scheme could be implemented and used to obtain the semi-classical limit and possibly better if we can exploit entangled photons.

In summary we establish that DE is a powerful machine-learning tool for devising adaptive quantum-enhanced metrology policies
and that our 
DE-based policy-devising algorithm significantly surpasses PSO for two important cases:
adaptive interferometric-phase estimation and estimating the bias of a quantum walker's coin.
This latter case entails using a $d$-ary decision tree where $d$ can be much greater than two,
and we show that a $d=4$ heuristic is effective even for large~$d$.
The power of the DE-based algorithm is evident in the fact that we double the number of particles solvable in a given computer time.
Given the $O(N^7)$ run-time cost of the algorithms, this means that we have an effective run-time speed-up of approximately $2^7$
over the previous best, namely the PSO-based algorithm.
Moreover, our new DE-based algorithm shows no sign of power-law deviation for double the number of particles compared to the
PSO-based algorithm, which means not only is there a run-time speed-up but also that the policies show improvement right up
to the data point for the largest particle number.
Finally we show that our adaptive quantum metrology policy-devising algorithm can be effected with current optical quantum-walk technology.
Policies for quantum metrology in the presence of phase noise and decoherence of the multi-photon state are known using PSO~\cite{HS11}, but DE algorithms for these conditions are a topic for future work.

We thank A. Hentschel and Y. R. Sanders for valuable discussions.
This research has been enabled by the use of computing resources provided by WestGrid and Compute/Calcul Canada
and was supported by CIFAR, NSERC, AITF and MPRIME.
MPL acknowledges financial support from La Caixa Foundation and AITF.
\bibliography{machine}

\end{document}